\documentclass{llncs}

\usepackage{hyperref}

\begin{document}

\title{Multi-Agent Programming Contest 2012 \\ --- \\ The Python-DTU Team}

\author{J{\o}rgen Villadsen \and Andreas Schmidt Jensen \and Mikko Berggren Ettienne \and Steen~Vester \and Kenneth Balsiger Andersen \and Andreas Fr{\o}sig}

\institute{Department of Informatics and Mathematical Modelling \\
Technical University of Denmark \\
Richard Petersens Plads, Building 321, DK-2800 Kongens Lyngby, Denmark}

\maketitle
\thispagestyle{plain} \pagestyle{plain}

\smallskip

\begin{abstract}
We provide a brief description of the Python-DTU system, including the overall design, the tools and the algorithms that we plan to use in the agent contest.

\

Updated 1 October 2012: Appendix with comments on the contest added.
\end{abstract}

\smallskip

\section*{Introduction}

\begin{enumerate}
\item
The name of our team is Python-DTU. We participated in the contest in 2009 and 2010 as the Jason-DTU team \cite{Boss+2010,Vester+2011},
where we used the Jason platform \cite{Bordini+2007}, but this year we use just the programming language Python
as we did in 2011 \cite{Ettienne+2012}.
\medskip
\item
The members of the team are as follows:
\begin{itemize}
\smallskip
\item J{\o}rgen Villadsen, PhD
\smallskip
\item Andreas Schmidt Jensen, PhD student
\smallskip
\item Mikko Berggren Ettienne, MSc student
\smallskip
\item Steen Vester, MSc student
\smallskip
\item Kenneth Balsiger Andersen, BSc student
\smallskip
\item Andreas Fr{\o}sig, BSc student
\smallskip
\end{itemize}
We are affiliated with DTU Informatics (short for Department of Informatics and Mathematical Modelling, Technical University of Denmark, and located in the greater Copenhagen area).
\medskip
\item
The main contact is associate professor J{\o}rgen Villadsen, DTU Informatics, email: \email{jv@imm.dtu.dk}
\medskip
\item
We expect that we will have invested approximately 200 man hours when the tournament starts.
\end{enumerate}

\section*{System Analysis and Design}

\begin{enumerate}
\item
The competition is built on the Java MASSim platform and the Java EISMASSim framework is distributed with the competition files.
This framework is based on EIS and abstracts the communication between the server and the agents to simple Java method calls and callbacks.
We decided to skip EISMASSim to instead follow a much cleaner Python-only implementation.
Even though some work was needed to implement the protocol specific parts which EISMASSim handled, this left us with a more flexible implementation
of which we have complete knowledge and control of every part of the implementation.
\medskip
\item
We do not use any existing multi-agent system methodology.
\medskip
\item
We do not plan to distribute the agents on several machines mainly for two reasons. Firstly, we had no need to, as we have enough computation power on a single machine to reason and send the action messages before the deadlines. Secondly, the shared data structure in our implementation would have to be replaced by a message server and a simple protocol. Due to limited time we have to prioritize differently.
\medskip
\item
We do not plan a solution with a centralization of coordination/information on a specific agent.
Rather we plan a decentralized solution where agents share percepts through through shared data structures and coordinated actions using distributed algorithms.
\medskip
\item
Our communication strategy is to share all new percepts to keep the agents internal world models identical. Furthermore our agreement based auction algorithm heavily relies on communication and is part of how agents decide on goals.
\medskip
\item
We hope to achieve the following properties when designing an algorithm to assign goals to agents:
\begin{enumerate}
\item The total benefit of the assigned goals should be as high as
  possible. Preferably optimal or close to it.
\item The running time of the algorithm should be fast, since we need
  to assign goals to agents at every time step in the competition and
  still have time left for other things such as environment
  perception, information sharing, etc.
\item The algorithm should be distributed between the agents
  resembling a true multi-agent system.
\item It should not be necessary for the agents to have the same
  beliefs about the state of the world in order to agree on an
  assignment.
\item The algorithm should be robust. If it is possible, our agents
  should be able to agree on an assignment even if some agents break
  down or some communication channels are broken.
\end{enumerate}
\medskip
\item
Each agent acts on its own behalf based on its local view of the world which is updated through percepts and is thus autonomous and reactive. This is implemented as an agent-control-loop in which the agents decide which actions to execute based on their current view of the world. When a repairer and a disabled agent moves towards each other the repairer decides and announces who should take the last step so they won't miss each other. This proactiveness is implemented by considering the current energy and the paths of the agents.
\end{enumerate}

\section*{Software Architecture}

\begin{enumerate}
\item
We do not use any multi-agent programming language.
We implement the multi-agent system using just the programming language Python.
\par\medskip
We choose Python as our programming language, as we think it has some advantages over for example Java, mainly in development speed/programmer effectiveness.
Some of the reasons being that Python in contrast to Java:
\begin{itemize}
\smallskip
\item is dynamically typed
\smallskip
\item is concise
\smallskip
\item is compact
\smallskip
\item supports multiple programming paradigms (object-oriented, imperative, functional)
\smallskip
\item is popular for scripting
\smallskip
\item does not need to be compiled before execution
\smallskip
\end{itemize}
\medskip
\item
We use Python 3.0 on Linux and Mac OS X as the development platforms and GEdit, Eclipse and TextMate as code editors/IDEs.
\medskip
\item
As the runtime platform for the competition we plan to use a suitable Linux system with Python 3.0.
\medskip
\item
Our implementation has mainly relied on custom best-first searches and a distributed auction-based agreement algorithm and a custom pathfinding algorithm tweaked for this domain.
\end{enumerate}

\

\section*{Acknowledgements}

Thanks to Per Friis for IT support.

\vfill

\begin{center}
More information about the Python-DTU team is available here:
\\[3ex]
\url{http://www.imm.dtu.dk/~jv/MAS}
\end{center}

\

\section*{Appendix}

The aim of the annual agent contest is to stimulate research in the area of multi-agent systems, to identify key problems and to collect suitable benchmarks.

\medskip

\noindent
The 2012 contest was organized by Tristan Behrens, J\"{u}rgen Dix, Michael K\"{o}ster and Federico Schlesinger at the Clausthal University of Technology, Germany.
The scenario and schedule were announced 20 February 2012 and the tournament took place 10-12 September 2012.

\medskip

\noindent
The 2012 winner was the Jason-UFSC team led by Jomi~Fred H\"{u}bner, Federal University of Santa Catarina, Brazil.
Like in 2011 we came in second. Both teams won all matches against the 5 other teams but we lost 1-2 against the winner.

\medskip

\noindent
The 5 other teams came from Brazil, China, Germany, Iran and USA.
In addition 2 teams from Germany and Ireland did not make it in the qualification phase where the stability of the teams had to be proved.

\medskip

\noindent
Further details are available here: \url{http://multiagentcontest.org}

\end{document}